\documentclass[journal,12pt,final]{IEEEtran}

\usepackage[utf8]{inputenc}
\usepackage[dvipsnames]{xcolor}
\usepackage{soul}
\usepackage{amsmath}
\usepackage{booktabs}
\usepackage{graphicx}
\usepackage{pgfplots}
\usepackage{pgfplotstable}
\usepgfplotslibrary{fillbetween}
\usepgfplotslibrary{groupplots}
\pgfplotsset{compat=1.17}
\usetikzlibrary{arrows.meta}
\usepackage{subfigure}
\usepackage[font=footnotesize]{caption}
\usepackage{paralist}
\usepackage{amssymb}
\usepackage[range-phrase=--,per-mode=symbol-or-fraction,range-units=single,list-units=single,detect-all]{siunitx}

\usepackage{cite} 
\usepackage{csquotes}

\usepackage{hyperref}
\usepackage[capitalise,noabbrev]{cleveref}
\Crefformat{figure}{#2Fig.~#1#3}
\Crefformat{equation}{#2Eq.~(#1)#3}

\ifCLASSINFOpdf
\else
\fi

\usepackage{acro}
\usepackage{mfirstuc}
\DeclareAcronym{MC}{short=MC, long=Molecular Communication}
\DeclareAcronym{IM}{short=IM, long=Information Molecules}

\usepackage{titlesec}
\titlespacing\section{0pt}{3pt plus 1pt minus 1pt}{2pt plus 1pt minus 1pt}
\titlespacing\subsection{0pt}{2pt plus 1pt minus 2pt}{2pt plus 0pt minus 1pt}

\IEEEoverridecommandlockouts 

\newcommand\hide[1]{}

\newcommand{\bitem}{\begin{itemize}}
\newcommand{\eitem}{\end{itemize}}
\newcommand{\benum}{\begin{enumerate}}
\newcommand{\eenum}{\end{enumerate}}
\newcommand{\bcitem}{\begin{compactitem}}
\newcommand{\ecitem}{\end{compactitem}}
\newcommand{\bcenum}{\begin{compactenum}}
\newcommand{\ecenum}{\end{compactenum}}

\begin{document}

\setlength{\skip\footins}{10pt}

\title{The Shared Prosperity Internet \
\vspace{-0.3cm}
}
\author{
    \IEEEauthorblockN{
        Juan A. Cabrera\IEEEauthorrefmark{1}\IEEEauthorrefmark{3},
        Pit Hofmann\IEEEauthorrefmark{1}\IEEEauthorrefmark{3},
        Jonas Schulz\IEEEauthorrefmark{1},
        Frederic Benken\IEEEauthorrefmark{1}\IEEEauthorrefmark{3},
        Hrjehor Mark\IEEEauthorrefmark{1}\IEEEauthorrefmark{3},\\
        Giang T. Nguyen\IEEEauthorrefmark{4}\IEEEauthorrefmark{3},
        Holger Boche\IEEEauthorrefmark{2}\IEEEauthorrefmark{3}, and
        Frank H.\,P. Fitzek\IEEEauthorrefmark{1}\IEEEauthorrefmark{3}\\
    }
    \IEEEauthorblockA{
        \IEEEauthorrefmark{1}Deutsche Telekom Chair of Communication Networks, TU Dresden, Germany\\
        \IEEEauthorrefmark{4}Haptic Communication Systems, TU Dresden, Germany\\
        \IEEEauthorrefmark{2}Chair of Theoretical Information Technology, TU Munich, Germany\\
        \IEEEauthorrefmark{3}Centre for Tactile Internet with Human-in-the-Loop (CeTI), Dresden, Germany\\
        \texttt{\{first\_name.last\_name\}@tu-dresden.de}; \texttt{boche@tum.de}
    }
    \thanks{The authors acknowledge the financial support by the German Research Foundation (DFG) as part of Germany's Excellence Strategy -- EXC 2050/2 -- Cluster of Excellence ``Centre for Tactile Internet with Human-in-the-Loop'' (CeTI) of TUD Dresden University of Technology under project ID 390696704, and by the Federal Ministry of Research, Technology and Space (BMFTR) of Germany in the program of ``Souverän. Digital. Vernetzt.'' Joint project 6G-life, grant numbers 16KIS2413K and 16KIS2414.}
    \thanks{The authors acknowledge the use of AI tools to create certain visual elements. Fig.~\ref{fig:challenges} includes waveforms generated with Midjourney and refined in Vizcom, color-adjusted in Photoshop, and annotated in Figma. Fig.~\ref{fig:use_case} includes three concept images generated with Midjourney (``Remote teaching for pupils'' (left), ``Remote teaching of cyber‑physical systems or robots'' (middle), ``Care for elderly people'' (right)); the right image was further edited in Photoshop to add the hologram.}
    \thanks{This work is a preprint. It has been submitted to the IEEE for possible publication.} 
    }

\maketitle
\begin{abstract}
The Shared Prosperity Internet (SPI) is a network–computing architecture that makes the benefits of automation and Artificial Intelligence (AI) broadly accessible to the society. To ground its design, this paper maps the physical constraints of Shannon, Landauer, Turing, and Einstein to three design principles: trustworthiness, sustainability, and technological sovereignty, and maps them into three technical pillars: $i)$ post-Shannon, goal-oriented communication that transmits only what the task requires; $ii)$ anticipatory decision-making (``negative latency'') with confidence-bounded pre-action and correction; and $iii)$ beyond-digital computing that selects energy-optimal substrates under deadline and computability constraints. The SPI is grounded in three societal use cases: remote teaching for pupils, remote teaching of robots and cyber-physical systems, and elder care. Furthermore, this paper defines measurable outcomes for an SPI, including latency decomposition, bits per event, energy/CO$_2$ per task, safety/privacy indicators, and robustness.
\end{abstract}

\IEEEpeerreviewmaketitle

\section{Introduction}
\begin{figure*}[!t]
  \centering
  \includegraphics[width=\textwidth]{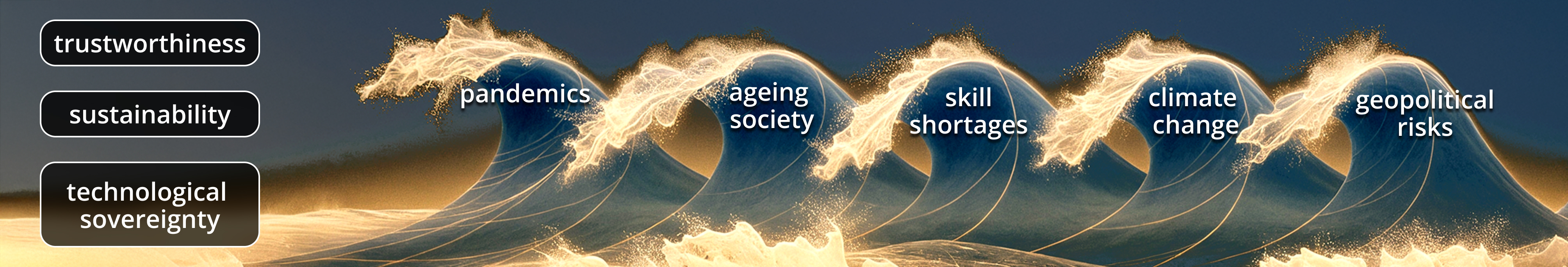} 
  \caption{The five societal challenges.}
  \label{fig:challenges}
  \vspace{-0.5cm}
\end{figure*}

In 2024, the Nobel Prize in Economic Sciences was awarded to D. Acemoglu, S. Johnson, and J. A. Robinson for their work showing how political institutions shape long-run prosperity. The committee’s scientific background summarises a central implication: democratic, inclusive institutions are a robust pathway to economic development, and more inclusive governments promote economic development~\cite{KVA2024EconomicSciencesScientificBackground}. Building on this insight, this paper proposes the Shared Prosperity Internet (SPI) as a digital institutional design. If inclusive governance promotes development, then an internet that embeds inclusivity by design should also broaden the benefits of technological progress. The idea of an SPI is simple: increases in prosperity generated by robotics, communication networks, or algorithms must benefit society as a whole, not just a small elite or individual companies. It is not about egalitarianism or state control, but about making technology accessible to everyone and not leaving it solely in the hands of oligarchs, monopolies, or despots. Evidence shows that societies are more successful and stable when everyone, from the smallest craftsmen in their workshops to the largest production lines in industrial parks, utilizes technological and economic advances. A good example of the successful implementation of shared prosperity in the technological domain is the mobile communication system. When fishermen in Kerala, India, adopted mobile phones, the price dispersion fell, and the amount of spoilage dropped, as they were able to navigate to the best-paying markets~\cite{Jensen2007}. Likewise, the internet was initially designed as a decentralized technology for people. Still, it is now becoming centralized in the hands of a select few, such as AI giants and social networks. The SPI is advocated, which returns power to the people by design. This is achieved by enabling humans and machines to collaborate across distances in both physical and virtual worlds. It goes far beyond state-of-the-art technology by drastically improving trustworthiness, sustainability, and technological sovereignty for communication and computing paradigms. This approach will introduce new applications in healthcare, caregiving, education, and work, addressing five key societal challenges: pandemics, aging society, skills shortages, climate change, and geopolitical risks, shown in Fig~\ref{fig:challenges}.

The aim of the SPI to tackle these key challenges imposes unprecedented demands on communication networks and computing platforms in terms of energy consumption, capacity, computability, and latency. However, attempting to satisfy these demands through a linear evolutionary process that encompasses classical communication and computing approaches is unlikely to succeed~\cite{Fettweis2014}.
Fundamental physical limitations, as defined by Shannon, Landauer, Turing, and Einstein, necessitate the development of new concepts and theories.

\section{The Fundamental Physical Limitations}

Deploying the SPI on a vast scale will face challenges, as described in this section, imposed by four fundamental physical constraints that have been known since the latter half of the 20th century. Shannon, Landauer, Turing, and Einstein offered insights into the physical limitations inherent in their theories. Hence, satisfying the demands of our envisioned SPI requires new paradigms for communication and computing systems that allow us to circumvent these limitations.

\subsection{Landauer's Limit}
Future communication systems will produce information at unprecedented scales. The infrastructure must transport and process every bit of information. However, one theory from the last century, proposed by Landauer~\cite{Landauer1961Irreversibility} and experimentally verified~\cite{Berut2012LandauerVerification}, holds that information is physical. This implies that every irreversible bit operation, e.g., an AND gate, incurs a theoretical lower bound of unavoidable energy cost of \num{3e-21} Joules at room temperature. Current processors consume much more energy than that, despite continuous efficiency improvements. However, Landauer’s principle limits the efficiency of conventional processors. As shown in~\cite{Ang2021DecadalPlanSRC}, if no disruptive research is carried out, the demand for Information and Communication Technology (ICT) could match global energy production by 2040. Landauer’s limit implies that scaling future communication systems cannot rely solely on linear evolutionary solutions. To overcome this energy barrier, disruptive approaches need to be investigated that either drastically reduce the amount of generated data (post-Shannon paradigms) or radically change how information is processed (analog or biological computing).

\subsection{Shannon's Limit}
Shannon’s limit specifies the maximum capacity of a communication channel at a given bandwidth and signal-to-noise ratio. However, this limit is defined for a specific communication task: the reproduction of symbols produced at one location at another. Shannon stated~\cite{Shannon1948MathematicalTheory} that the meaning and the intention of the conveyed symbols are outside the engineering problem. Only with such a general approach to communication was it possible to build the information age. Although Shannon’s limit may have been sufficient in the past decades, an internet that truly increases shared prosperity necessitates disruptive new approaches to communication. Post-Shannon paradigms promise to be a solution to the data flood: By incorporating the semantics and goals of communication as design paradigms of the systems, it is possible to drastically reduce traffic requirements. For example, it has been demonstrated that message identification can reduce data traffic exponentially~\cite{vonLengerke2025IDviaChannelsTutorial}.

\subsection{Turing's Limit}
In today’s digital world, little to no thought is given to the limits of digital computers. Their use is so widespread and their applications so vast that one tends to forget there is an inherent limitation to what they can do. The ideal digital computer is the Turing machine~\cite{Turing1937ComputableNumbers}, i.e., a machine capable of computing any computable function. However, the set of actually computable functions is relatively small. This limitation implies that there is a large (uncountably infinite) set of functions that cannot be computed by digital means~\cite{Boche2023DigitalTwinOptimization,Bock2023VirtualTwinNetworking}. This raises the question: Can all functions required by an SPI be computed digitally? Previous research has demonstrated that it is not possible to compute optimal communication codes for arbitrary channels~\cite{Boche2021TuringMeetsShannonICC} nor to compute whether an intelligent jammer attack is targeting a communication system~\cite{Boche2020DoSDetectability} using Turing machines. Many more essential functions for an SPI are non-computable. This leads to a fundamental research question: If a computation by digital means is not possible, can alternative computing models, such as a Blum-Shub-Smale machine, be potential solutions?

\subsection{Einstein's Limit}
The SPI faces a hard physical wall: nothing can propagate faster than the speed of light. This implies an irreducible end-to-end latency for any interaction that depends on information exchange over distance. For many SPI use cases, such as teleoperation with haptics, remote healthcare, or distributed industrial control, this propagation latency is already longer than what humans perceive as ``instantaneous''. Classical network design can only push signals closer to the user; it cannot eliminate the propagation delay itself. The only viable path is anticipatory communication and control. By predicting user intent (e.g., gaze shifts, grasp targets, motor commands) and machine intent (e.g., short-horizon actuation trajectories) before those intents are fully expressed, the SPI can act locally on forecasts of the future rather than reactively on delayed measurements. This is the concept of ``negative latency''~\cite{Schulz2024NegativeLatency}, in which the effective response time appears to be less than zero because the system has already prepared the next state and only needs to correct deviations when the real signal arrives. In other words, Einstein’s limit cannot be broken, but it can still be functionally bypassed at the application level by embedding prediction, shared anticipatory state, and confidence-bounded pre-action directly into the infrastructure.

\begin{figure*}[!t]
  \centering
  \includegraphics[width=\textwidth]{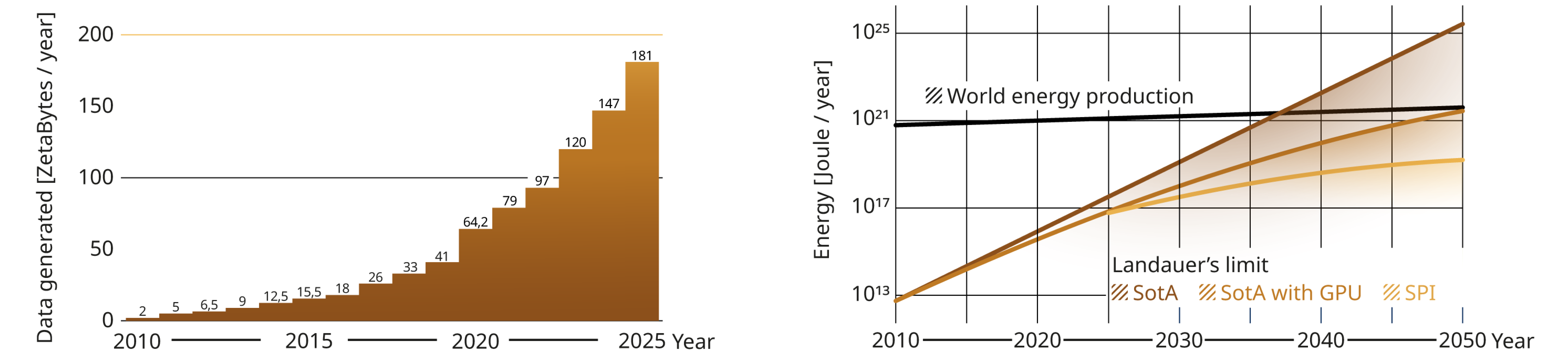} 
  \caption{Worldwide data generation over the years (left); worldwide energy production and consumption for state of the art (SotA), SotA with acceleration boards like GPUs, and the SPI approach (right); data based on~\cite{Ang2021DecadalPlanSRC}.}
  \label{fig:energy-traffic}
  \vspace{-0.5cm}
\end{figure*}

\section{Design Principles of the SPI}
Realizing the SPI concept will enable us to: $i)$ reduce physical contact during pandemics, $ii)$ enhance the interaction between humans, machines, and digital worlds through additional senses such as touch, which will significantly benefit elder care in aging societies, and $iii)$ enable experts around the globe to tele-operate with cyber-physical systems such as robots, thereby reducing the need for travel. Thus, the physical limitations from the aforementioned section are mapped onto three design principles: $(i)$ trustworthiness, $(ii)$ sustainability, and $(iii)$ technological sovereignty.

\subsection{Trustworthiness}
Latency is crucial for enabling trustworthy human–machine interaction because delayed system responses disrupt information flow, reduce perceived reliability, and impair the user’s sense of control and predictability. Latency comprises several components, including message delay (Shannon), computing delay (Turing), and propagation delay (Einstein). An SPI must address message and propagation latency, significantly reducing data volume by applying post-Shannon paradigms and extending the concept of negative latency, respectively. Disruptive post-Shannon paradigms address the challenge of generating information beyond Shannon’s limit by incorporating the semantics and goals of communication into the design. The concept of negative latency uses computing to predict or to learn about future behavior. This concept was introduced in~\cite{Schulz2024NegativeLatency}, demonstrating negative delays of up to 800\,ms. For an SPI, this approach must be extended to a wide range of use cases and increase negative latency by seconds. Furthermore, new computing platforms described below will enhance trustworthiness by reducing computing delays and addressing the computability issue identified by Turing. Thus making computation more reliable.

\subsection{Sustainability}
If the amount of generated information continues to increase (see Fig.~\ref{fig:energy-traffic} left), the energy consumption of the ICT sector will account for all worldwide energy production by 2040~\cite{Ang2021DecadalPlanSRC} (see Fig.~\ref{fig:energy-traffic} right). 
Thereby, latency, energy consumption, and computation are inherently coupled. An SPI will require task-oriented frameworks that, given a specified task, select the most suitable computing substrate (digital, analog, biological, or quantum) and tune computing power (e.g., clock frequency) to meet the required latency and trustworthiness (precision of the output) while minimizing energy consumption.

\subsection{Technological Sovereignty}
An SPI aims to achieve sovereignty at the system level, enabling operators and communities to develop, assess, and modify critical communication systems in accordance with their own guidelines and priorities. The framework accelerates prototype development through open-reference stacks and reproducible test environments. It also guarantees transparent, verifiable data paths and control interfaces, while reducing vendor lock-in through open Application Programming Interfaces (APIs), portable codes and models, and compliance profiles tailored to different legal contexts.

\section{Three Technical Pillars for an SPI}

Three technical pillars are identified that will sustain an SPI: Pillar~1 rethinks communication paradigms to enable semantic and goal-oriented communication within the SPI. Pillar~2 exploits AI- and Machine Learning (ML)-based predictive approaches, allowing the SPI infrastructure to react to predicted human actions before sensor-actuator information is propagated over the network. Pillar~3 focuses on integrating low-energy, beyond-digital computing platforms into the SPI infrastructure. Pillars~1 and 2 aim to reduce end-to-end delay and energy consumption by shrinking the transmission time and converting part of the propagation time into negative latency. This shifts the workload to computation, which is precisely what Pillar~3 addresses with low-energy, beyond-digital computing to preserve and amplify these gains.
\begin{figure*}[!t]
  \centering
  \includegraphics[width=\textwidth]{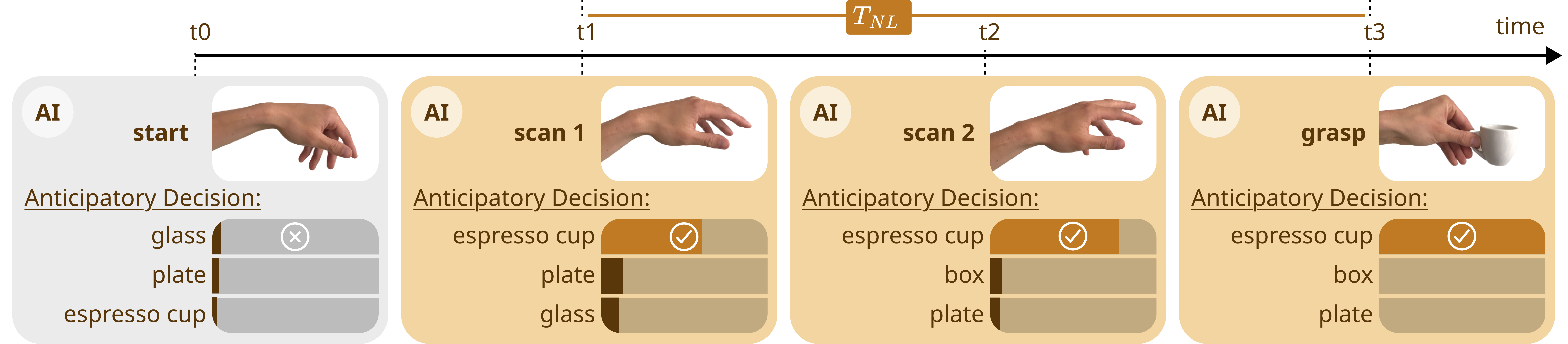} 
  \caption{Negative latency in human-machine interaction. As a person reaches to grasp an object, the system interprets muscle activity to detect intention signals. Negative latency ($T_\mathrm{NL}$) occurs when subsequent processes begin before the movement reaches its intended goal.}
  \label{fig:neglatency}
  \vspace{-0.5cm}
\end{figure*}

\subsection{Semantic- and Goal-Oriented Communication: Shannon's and Landauer's Limit}
To support the SPI, communication systems must address several key technical parameters that are interconnected and complex to optimize simultaneously, including latency, data rate, security, and massiveness, while also significantly reducing their energy consumption. Large numbers of sensor-driven human–machine interfaces require very high data rates, which will negatively impact latency. Latency arises from several reasons, such as medium access, transmission, computing, and propagation delays. The current approaches to 5G and 6G mobile communication systems primarily focus on increasing data rates and reducing channel access times. These linear evolutionary steps are insufficient for the envisioned SPI. Thus, groundbreaking research requires disruptive techniques: new information-theoretical approaches and quantum communication. Post-Shannon paradigms, such as semantic- and goal-oriented communication, can significantly reduce data rates, substantially decrease energy consumption, and dramatically improve latency for data transmission, storage, and computing. For the past few decades, the Shannon limit has been accepted as the ultimate bound on capacity. However, research has shown that semantic and goal-oriented communication enables a dramatic reduction in data traffic without compromising the success of the communication goal~\cite{vonLengerke2025IDviaChannelsTutorial}. Interestingly, post-Shannon approaches enable the use of resources that are useless in traditional transmissions. For example, it has been proven that access to Shared Randomness (SR) as a resource, i.e., the shared knowledge of the outcomes of a random variable, does not increase transmission capacity. However, if the goal is message identification (practical in event-driven communication), SR is a valuable resource that improves the identification capacity~\cite{vonLengerke2025IDviaChannelsTutorial}. This opens the door to novel approaches such as quantum communication, which enables quantum entanglement for SR and reduces communication latency with protocols like superdense coding~\cite{PhysRevLett.69.2881}. The latter allows sharing randomness in advance to increase the rate of communication when needed. Within a post-Shannon paradigm, SR generation is a novel communication task~\cite{Ezzine2025UniformCR} that can be applied to other post-Shannon tasks, as well as network coding and compressed sensing. An SPI must further explore the benefits of randomness and other previously discarded resources within these new communications paradigms.

\subsection{Anticipatory Communication \& Control: Einstein’s Limit}
To keep the propagation delay low, computing is placed within the network in a concept often referred to as the \emph{mobile edge cloud}. There are significant contributions to the field, including novel ideas such as the combination of quantum communication and computing, as mentioned above. The main idea is to mitigate virtual computing delays by performing quantum in-network computing operations directly on the physical link rather than at higher layers. However, the SPI must go beyond these improvements by targeting the fourth fundamental barrier: propagation delay. No signal, control loop, or perception–action cycle can propagate faster than the speed of light. For tightly interactive tasks such as teleoperation, assistive robotics, augmented and virtual reality with haptics, distributed industrial actuation, and cooperative manipulation, that delay is already unacceptable at human perception thresholds. The SPI cannot simply transmit faster; it must act earlier. Therefore, it must embed prediction into the infrastructure. Instead of waiting for sensor data to propagate and then reacting, it must forecast human and machine intent and pre-act locally. This extends the concept of negative latency: the effective response time at an endpoint becomes \emph{less than zero}, because the next state is prepared in advance and only corrected (not initiated) when the real signal arrives~\cite{Schulz2024NegativeLatency}; see Fig~\ref{fig:neglatency}. Negative latency was first introduced in cloud gaming, where the server predicts the next player input and renders frames ahead of time. However, it has been demonstrated at the motor-control level in human grasping, where predicted intent drives assistive actuation before the physical command is fully propagated~\cite{Schulz2024NegativeLatency}. An SPI must extend negative latency to the eye and brain (anticipating gaze shifts and intended motion before they occur) and to machine-to-machine interaction via predictive intent codecs that transmit short-term future behavior rather than current state. This anticipatory layer is what allows the SPI to maintain safe, coordinated action across distance despite the speed-of-light limitation.

\begin{figure*}[!t]
  \centering
  \includegraphics[width=\textwidth]{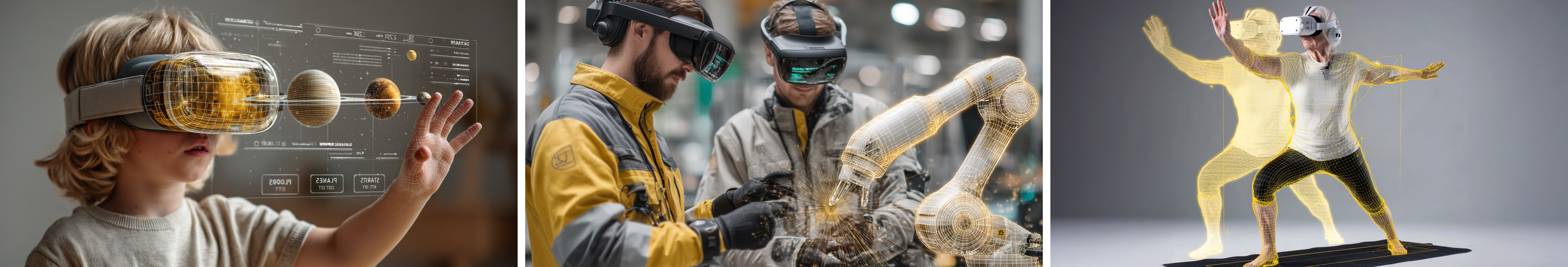} 
  \caption{SPI use cases: Remote teaching for pupils (left); Remote teaching of cyber-physical systems or robots (middle); Care for elderly people (right).}
  \label{fig:use_case}
  \vspace{-0.5cm}
\end{figure*}

\subsection{Beyond-Digital Computing: Landauer's and Turing's Limit}
Facing the questions: How can the energy required for the computations of an SPI be reduced, and is it possible to compute all its necessary tasks digitally? To this end, new computing hardware platforms need to be explored, and their integration in the SPI. Initially, to address the energy consumption of computing, small AI models on digital platforms serve as a starting point, such as TinyML, or by outsourcing AI complexity through decentralized approaches, such as EdgeML, or by using neuromorphic computing hardware in digital substrates. However, analog and biological computing systems must be investigated that promise a thousand times energy savings, and understand the fundamental connection between these computing platforms and the Landauer limit. Quantum computing is an exception here. The combination of quantum computing and communication offers an extreme solution for achieving very low latency, but this comes at the cost of increased energy consumption. This is intended for rare and extreme use cases, but not the everyday tasks of the SPI. In parallel, the computability of the SPI tasks must be addressed. Turing has already demonstrated that digital hardware is limited by highlighting the concept of uncomputable functions. Even though humans live in the analog world, communication and machines are typically designed to be digital (for noise reduction and general-purpose use). If the new tasks for an SPI can be calculated in the analog domain, there would still be the challenge of noise and integration within the infrastructure. Therefore, the research on the fundamental computability characteristics and practical implementation aspects of an SPI across neuromorphic, analog, and biological platforms needs to be extended. Regarding the trustworthiness of computing, this paper postulates that analog implementations could surpass conventional digital designs in specific metrics. Conversely, the trust properties of biological platforms remain largely unexplored and constitute an open research area for the SPI.

\section{Potential Use Cases}

This section introduces three potential use cases for an SPI, see~\cref{fig:use_case}. They are especially relevant to address the goal of shared prosperity. They include people of all ages. Use case one focuses on the teaching of young pupils, enhancing their learning experience to prepare them for the future. Use case two focuses on adults, production, and labor by providing communication networks that enable the teaching and use of robots and cyber-physical systems not only in large factories and production lines, but also in the hands of small crafters, such as bakers, tailors, and medium and small manufacturers in general. Use case three addresses aging societies by using technology to enhance care for the elderly, thus providing an enjoyable and dignified experience in their later years.

\subsection{Remote Teaching for Pupils}
Deliver rich, interactive classes using significantly less data and energy compared to today's video-first setups. SPI utilizes post-Shannon codecs (e.g., identification and functional compression) in combination with haptic feedback to transmit only the information necessary for the learning task. This operates over 5G/6G campus network platforms, using shared randomness (via wireless or quantum communication systems) to reduce control traffic while enhancing security and privacy.
\textbf{Metrics:} End-to-end latency decomposition (covering message, computing, and propagation times), data traffic, joules per hour of instruction, privacy and trust indicators derived from randomness services, and scenario-level CO$_2$ estimates.

\subsection{Remote teaching of Robots}
Enable experts to teach and teleoperate cyber-physical systems remotely with negative-latency responses that feel immediate. SPI fuses anticipatory human-machine and machine-machine communication models with in-network computing. Post-Shannon protocols confirm model alignment of digital twins and send only corrective deltas. Testbeds utilize robotic labs, campus 5G, and Software-Defined Radios for tight control loops, as well as neuromorphic, analog, and biological computing to minimize deadline misses and energy consumption. Success ratios, task completion times, safety margins, rate reductions, energy consumption, and CO$_2$ emissions are benchmarked.
\textbf{Metrics:} time-to-first-correct action under partial information, correction rate after ground-truth arrival, network-induced vs. compute-induced latency, joules-per-task, and safety Key Performance Indicators (KPI) violations.

\subsection{Care for Elderly People}
This use case envisions proactive, privacy-preserving assistance (e.g., fall detection and intent-aware interaction) without the need for constant high-bit-rate sensing. SPI applies goal-oriented communication to event-driven sensing and anticipatory human-machine communication, enabling action on early biosignals while escalating fidelity only when confidence is warranted. A task-aware optimized computing substrate selector chooses digital, analog, neuromorphic, or biohybrid modules to meet deadlines with minimal energy; shared randomness hardens security and coordination.
\textbf{Metrics:} detection and false-alarm trade-off at fixed latency budgets, bits per detected event, joules-per-decision, privacy/trust, and CO$_2$ per patient-hour.

\section{Conclusion}
The SPI reframes networks as task-aware systems that respect four physical limits while delivering measurable gains in latency, energy, computability, and trust. This paper argued for three technical pillars: post-Shannon, goal-oriented communication; anticipatory decision-making; and substrate-aware computing, and demonstrated how they are composed through an orchestration layer that selects where and how to act. Thus, three potential cases for shared prosperity that provide concrete targets (e.g., bits per event, deadline miss rate, joules per task) have been presented. The claim is not that ``everything is solved'', but that SPI offers a disciplined way to trade bits, joules, and trust, so the broader society can see what they gain, what they pay, and where the limits bind.

\bibliographystyle{IEEEtran}
\bibliography{IEEEabrv,references}
\vspace{-1.3cm}

\begin{IEEEbiographynophoto}
{Juan A. Cabrera}
obtained his Dr.-Ing. degree at TU Dresden, Germany, in 2022, and his master's degree at Aalborg University, Denmark, in 2015. 
He currently works at the Deutsche Telekom Chair of Communication Networks at TU Dresden, where he leads a research group on semantic and goal-oriented communications. 
\end{IEEEbiographynophoto}
\vspace{-0.3cm}

\vskip -2\baselineskip plus -1fil
\begin{IEEEbiographynophoto}
{Pit Hofmann}~received the Dr.-Ing. degree from TUD Dresden University of Technology, Germany, in 2026. He is currently a Junior Research Group Leader with the Deutsche Telekom Chair of Communication Networks at TUD, and the Centre for Tactile Internet with Human-in-the-Loop (CeTI), Dresden, Germany, where he researches on molecular communication and biological computing.
\end{IEEEbiographynophoto}
\vspace{-0.3cm}

\vskip -2\baselineskip plus -1fil
\begin{IEEEbiographynophoto}
{Jonas Schulz}~is a PhD candidate at the Deutsche Telekom Chair of Communication Networks at TUD Dresden University of Technology, Germany. He received his Dipl.-Ing. degree in Electrical Engineering from TUD in 2022. His research focuses on human-centered communication systems, tactile Internet applications, and predictive human–machine interaction.
\end{IEEEbiographynophoto}
\vspace{-0.3cm}

\vskip -2\baselineskip plus -1fil
\begin{IEEEbiographynophoto}
{Frederic Benken}
is a PhD candidate at the Deutsche Telekom Chair of Communication Networks at TUD Dresden University of Technology, Germany. He received his Dipl.-Ing. degree in Mechanical Engineering from TUD in 2023. His research interests include human-centered design, user interface development, and the integration of virtual and physical prototyping in engineering.
\end{IEEEbiographynophoto}
\vspace{-0.3cm}

\vskip -2\baselineskip plus -1fil
\begin{IEEEbiographynophoto}
{Hrjehor Mark}
received the Dipl.-Ing. degree in electrical engineering from TUD Dresden University of Technology, Germany, in 1994. He has participated in numerous research projects in the field of ICT, spanning both industry and academia, with a strong focus on European contexts. He is currently serving as Program Office Director for the Cluster of Excellence CeTI, Dresden, Germany.
\end{IEEEbiographynophoto}
\vspace{-0.3cm}

\vskip -2\baselineskip plus -1fil
\begin{IEEEbiographynophoto}
{Giang T. Nguyen}
is an Assistant Professor heading the Haptic Communication Systems research group at the Cluster of Excellence CeTI and the Faculty of Electrical and Computer Engineering, TUD Dresden University of Technology, Germany. He received a Ph.D. degree in Computer Science from TUD in 2016. Earlier, he received the M.Eng. degree at the Asian Institute of Technology (AIT), Thailand.
\end{IEEEbiographynophoto}
\vspace{-0.3cm}

\vskip -2\baselineskip plus -1fil
\begin{IEEEbiographynophoto}
{Holger Boche} received the Dr.-Ing. degree in electrical engineering and the Dr. rer. nat. degree in mathematics. He was the Director of the Fraunhofer Institute for Telecommunications, Heinrich-Hertz-Institut, Berlin, Germany.
He is a Full Professor at the TU Munich, Germany, where he leads the Chair of Theoretical Information Technology. 
\end{IEEEbiographynophoto}
\vspace{-0.3cm}

\vskip -2\baselineskip plus -1fil
\begin{IEEEbiographynophoto}
{Frank H. P. Fitzek} is a professor and head of the Deutsche Telekom Chair of Communication Networks at TUD Dresden University of Technology, Germany.
He serves as the spokesman for the DFG Cluster of Excellence CeTI and the 6G-life hub in Germany.
He received his Dipl.-Ing. degree in electrical engineering from the Aachen University of Technology RWTH, Germany, in 1997 and his Dr.-Ing. degree from the TU Berlin, Germany, in 2002.
\end{IEEEbiographynophoto}

\end{document}